\pdfoutput=1
\documentclass{JINST}
\usepackage{subfigure}

\title{Characterization of Thin Pixel Sensor Modules Interconnected with SLID Technology Irradiated to a Fluence of 2$\cdot$10$^{15}$\,n$_{\mathrm{eq}}$/cm$^2$.}

\author{P. Weigell$^a$\thanks{Corresponding
author.}, L. Andricek$^{a,b}$, M. Beimforde$^a$, A. Macchiolo$^a$, H.-G. Moser$^{a,b}$, R. Nisius$^a$~ and R.-H. Richter$^{a,b}$\\
\llap{$^a$}Max-Planck-Institut f\"ur Physik,\\
  F\"ohringer Ring 6, 80805 M\"unchen, Germany\\
\llap{$^b$}Max-Planck-Institut Halbleiterlabor,\\
  Otto-Hahn-Ring 6, 81739 M\"unchen, Germany\\
  E-mail: \email{Philipp.Weigell@mpp.mpg.de}}

\abstract{A new module concept for future ATLAS pixel detector upgrades is presented, where thin n-in-p silicon sensors are connected to the front-end chip exploiting the novel Solid Liquid Interdiffusion technique (SLID) and the signals are read out via Inter Chip Vias (ICV) etched through the front-end. This should serve as a proof of principle for future four-side buttable pixel assemblies for the ATLAS upgrades, without the cantilever presently needed in the chip for the wire bonding.

The SLID interconnection, developed by the Fraunhofer EMFT, is a possible alternative to the standard bump-bonding. It is characterized by a very thin eutectic Cu-Sn alloy and allows for stacking of different layers of chips on top of the first one, without destroying the pre-existing bonds. This paves the way for vertical integration technologies. 

Results of the characterization of the first pixel modules interconnected through SLID as well as of one sample irradiated to $2\cdot10^{15}$\,\neqcm{} are discussed.

Additionally, the etching of ICV into the front-end wafers was started. ICVs will be used to route the signals vertically through the front-end chip, to newly created pads on the backside. In the EMFT approach the chip wafer is thinned to (50--60)\,$\mu$m.}


\newcommand{\neqcm}{\ensuremath{\mathrm{n}_{\mathrm{eq}}/\mathrm{cm}^2}}

\newcommand{\Sr}{\ensuremath{^{90}}Sr}
\newcommand{\Am}{\ensuremath{^{241}}Am}

\begin{document}
\section{Introduction}
\label{sec:introduction}
To reach even higher luminosities the LHC accelerator will be upgraded in two steps \cite{SLHC}. Starting in 2016, the aim is to reach a luminosity of around (2-3)$\cdot$10$^{34}$\,cm$^{-2}$s$^{-1}$ by some comparatively small upgrades to the machine itself and improvements to the pre-accelerators. Presumably after 2020, a major upgrade will increase the luminosity further to at least 5$\cdot$10$^{34}$\,cm$^{-2}$s$^{-1}$. In this scenario the innermost layers of the ATLAS vertex detector system will have to sustain very high integrated fluences of more than 10$^{16}$\,\neqcm{} (1\,MeV neutron equivalent).

Shortly before the start of the LHC upgrades, in the year 2013 a first upgrade of the ATLAS pixel detector, the Insertable B-Layer (IBL) \cite{IBL-TDR}, consisting of an additional pixel layer mounted onto a new beam pipe of smaller radius is envisaged. Another upgrade, called NewPix, is under discussion for the year 2018, where, depending on the operational performance, a replacement of the entire pixel detector is maybe needed. The pixel module concept investigated in this paper targets for this latter and following upgrades. 

\section{The Module Concept}
The module concept builds upon four novel technologies in the field of pixel detectors. N-in-p pixel sensors are thinned using a Max-Planck-Institut Halbleiterlabor (MPI-HLL) developed process \cite{Lacithin} and connected via the EMFT Solid Liquid Inter-Diffusion (SLID) technology to the read-out electronics, where the signals are routed via Inter-Chip-Vias (ICV). This approach has several advantages. The n-in-p bulk material allows for single sided processing of wafers resulting in a cost reduction, which is of special importance for the large volumes needed in future pixel detector upgrades at large radii. In addition, the material is comparable in radiation hardness to the presently used n-in-n bulk material \cite{Weigell2011,Annatipp}. Exploiting thinner sensors reduce on the one hand the material budget and therefore multiple scattering. On the other hand they exhibit higher electric fields than thicker devices, which leads to a higher charge collection efficiency after the high radiation doses expected in the future environment of the LHC. In this environment read-out electronics have to be capable of operation with the lower signal charges generated. First results for the new ATLAS read-out chip FE-I4 \cite{GarciaSciveres2010}, developed for the IBL-upgrade indicate that thresholds as low as 1.6\,ke are operable \cite{MaltePSD9}, which would be well sufficient for the sensor thicknesses around 75\,$\mu$m presented in these proceedings. For the present read-out chip FE-I3 \cite{Peric2006178}, used in the R\&D programme at the moment, thresholds down to 3.2\,ke are generally achievable. For some single chip modules (SCM) lower thresholds down to (2.0 -- 2.5)\,ke are achievable. 

N-in-p pixel sensors with active thicknesses of 75\,$\mu$m and 150\,$\mu$m have been produced from wafers of standard thickness using the  MPI-HLL thinning process. The pre-irradiation characterization of these sensors shows a very good device yield and high break down voltages. After irradiations up to a fluence of $10^{16}$\,n$_{\mathrm{eq}}$cm$^{-2}$ charge collection efficiency measurements yield charge collection efficiencies close to unity for the 75\,$\mu$m thick sensor \cite{MichaelTWEPP}.

\subsection{SLID Interconnection}
The SLID interconnection, developed by the Fraunhofer EMFT \cite{KlumppSLID}, is a possible alternative to the standard bump-bonding. It is characterized by a very thin eutectic Cu-Sn alloy and allows for stacking of layers of chips on top of the first one, without destroying the pre-existing bonds. This paves the way for exploiting vertical integration technologies. To prevent contamination with copper, the sensors as well as the read-out chips are covered with a 100\,nm thick TiW diffusion barrier \cite{Macchiolo2008229}. On this layer a 5\,$\mu$m thick copper layer is electroplated. On one side only a 3\,$\mu$m thick tin layer is added. After alignment of the sensor with respect to the read-out chip a pressure of several atmospheres and a temperature between 240\,$^{\circ}$C and 320\,$^{\circ}$C is applied to form the Cu-Sn alloy which forms the connection. This alloy exhibits a melting point around 600\,$^{\circ}$C and thus does not dissolve if additional layers are added with consecutive SLID interconnection. Another advantage of the SLID interconnection is that the reflow step, which is needed for standard bump bonding is not needed, resulting in less process steps and eventually lower cost. 
 
The process can be carried out in a wafer-to-wafer or chip-to-wafer approach, where in the latter case a handle wafer is used. Results presented here were obtained with SCM connected in the chip-to-wafer process. In this process the population of the handle wafer is challenging since an alignment of the chips better than 20\,$\mu$m is needed. Five out of ten SCMs were successfully produced.

\subsubsection{SLID SCM Characterization}
The current read-out chip of the ATLAS experiment (FE-I3) is used. The data were gathered using the USBPix system \cite{USBPix}. Source measurements were conducted using a \Sr{} and an \Am{} source. For the source measurements with \Sr{} a scintillator coupled to a photo-multiplier is used as external trigger and for the \Am{} measurements the internal trigger of the read-out chip is used. 

\subsubsection{Measurement with the Non-irradiated SCMs}
All five SCMs exhibit leakage currents below 0.1\,$\mu$A and breakdown voltages beyond 120\,V, which is well above full depletion occurring around 40\,V. 

The charge collection performance is within uncertainties the same over the five devices. The conversion from time-over-threshold to charge is affected by an uncertainty of 20\,\%. This is true as well for the threshold value of the chip resulting from the tuning procedure. Another uncertainty arises from the fact that charges below threshold are lost. This effect is estimated to be at the level of a few percent due to the small thickness of the devices and at most 8$^{\circ}$ of inclination angle for the tracks. For \Am{} measurements threshold effects are only important for the lower part of the spectrum, given the signal to threshold ratio.

\begin{figure}[h!t]
\centering
\subfigure[]{
\includegraphics[scale=0.35]{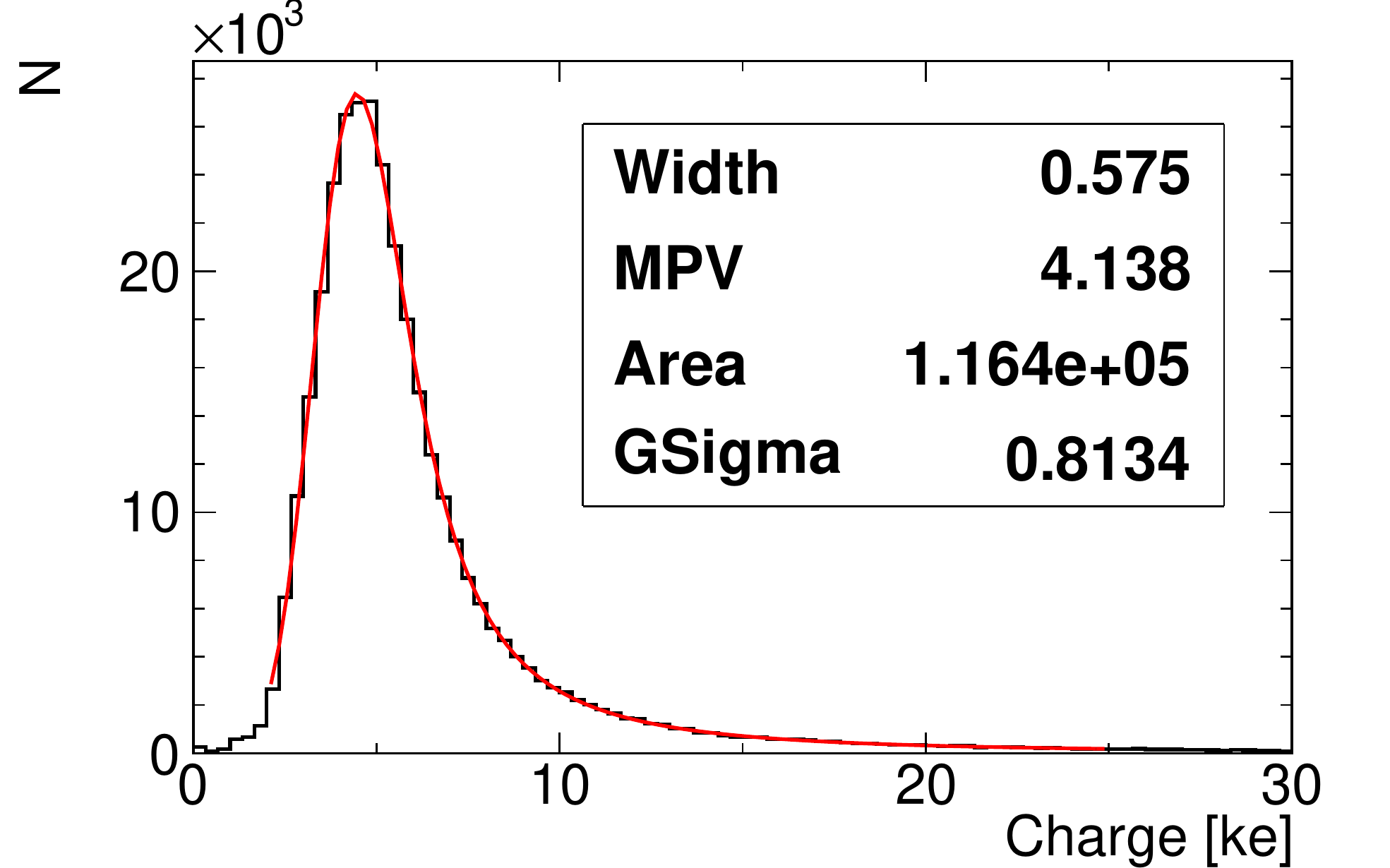}
\label{fig:SLID9Sr90}
}
\subfigure[]{
\includegraphics[scale=0.35]{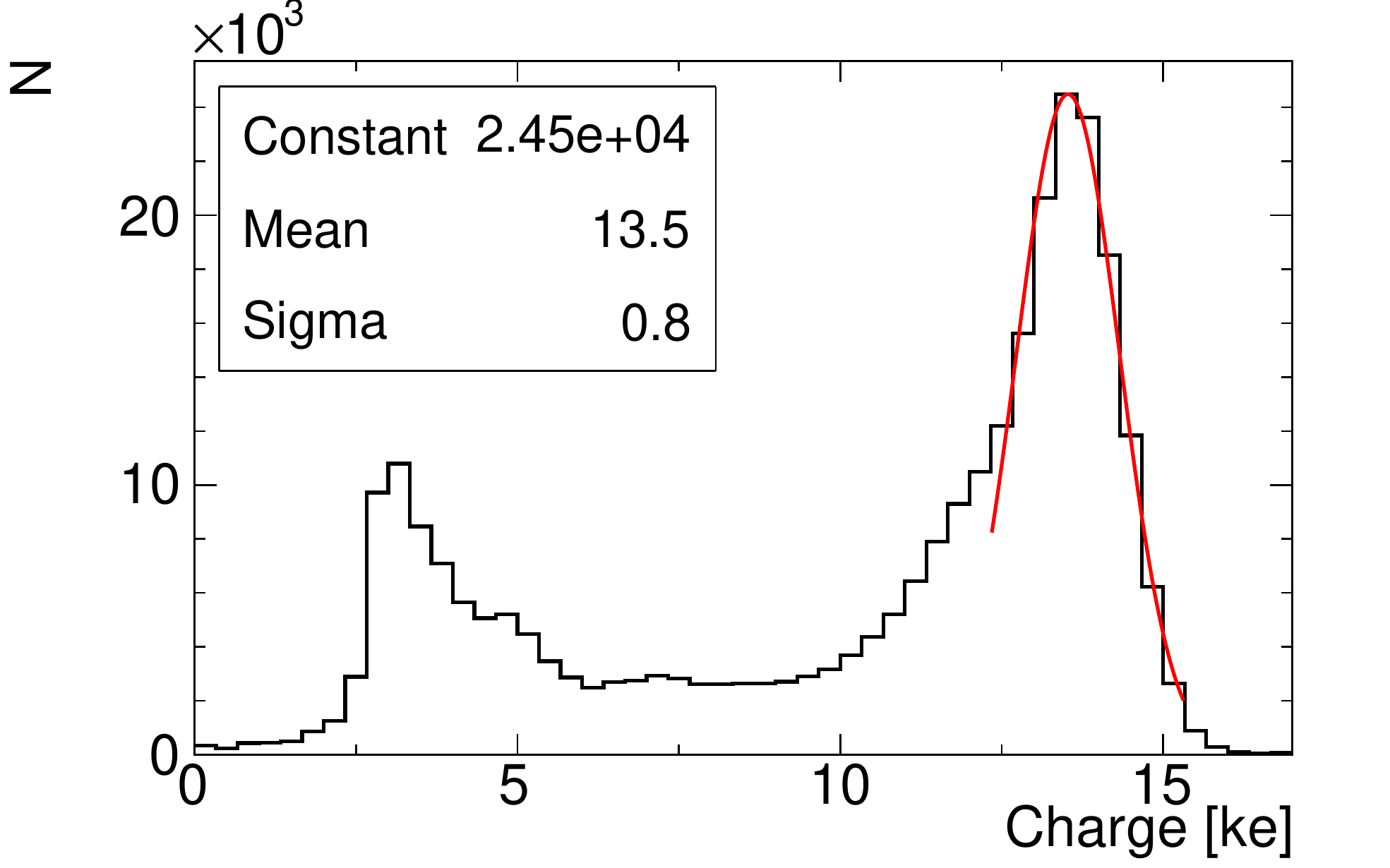}
\label{fig:SLID9Am241}
}
\caption{Spectrum for (a) a \Sr{} source measurement and (b) a \Am{} source measurement before irradiation. The \Sr{}-spectrum is fitted with a Landau convoluted with a Gaussian. The \Am{}-spectrum is fitted with a Gaussian.}
\label{fig:Charge_SourceScan}
\end{figure}

Example spectra from one device at a bias voltage of 55\,V are shown in Fig.\,\ref{fig:Charge_SourceScan}. The threshold was tuned to $2.8$\,ke for these measurements and had a dispersion of $31$\,e over the chip.
A Landau distribution convoluted with a Gaussian gives a good description of the \Sr{} measurement (cf.\,Fig.\,\ref{fig:SLID9Sr90}) and yields a most probable charge value (MPV) of $(4.14\pm0.8)$\,ke, which is within expectations for devices with an active thickness of 75\,$\mu$m \cite{Bichsel1988}.

In the spectrum of \Am{} shown in \ref{fig:SLID9Am241} the 59\,keV $\gamma$-line is seen at $(13.5\pm2.7)$\,ke, which is slightly below the expected value of 16.3\,ke. At 4\,ke  an additional peak corresponding to several lines of the \Am{} spectrum between 10 and 26\,ke convoluted with noise is seen. Most of this peak is cut off since it is below threshold. This performance, taking the different active thicknesses into account, matches the performance of standard ATLAS pixel modules in terms of charge collection and threshold dispersion.

\subsubsection{Measurement with a Neutron Irradiated SCM}
\begin{figure}[h!t]
\centering
\subfigure[]{
\includegraphics[scale=0.35]{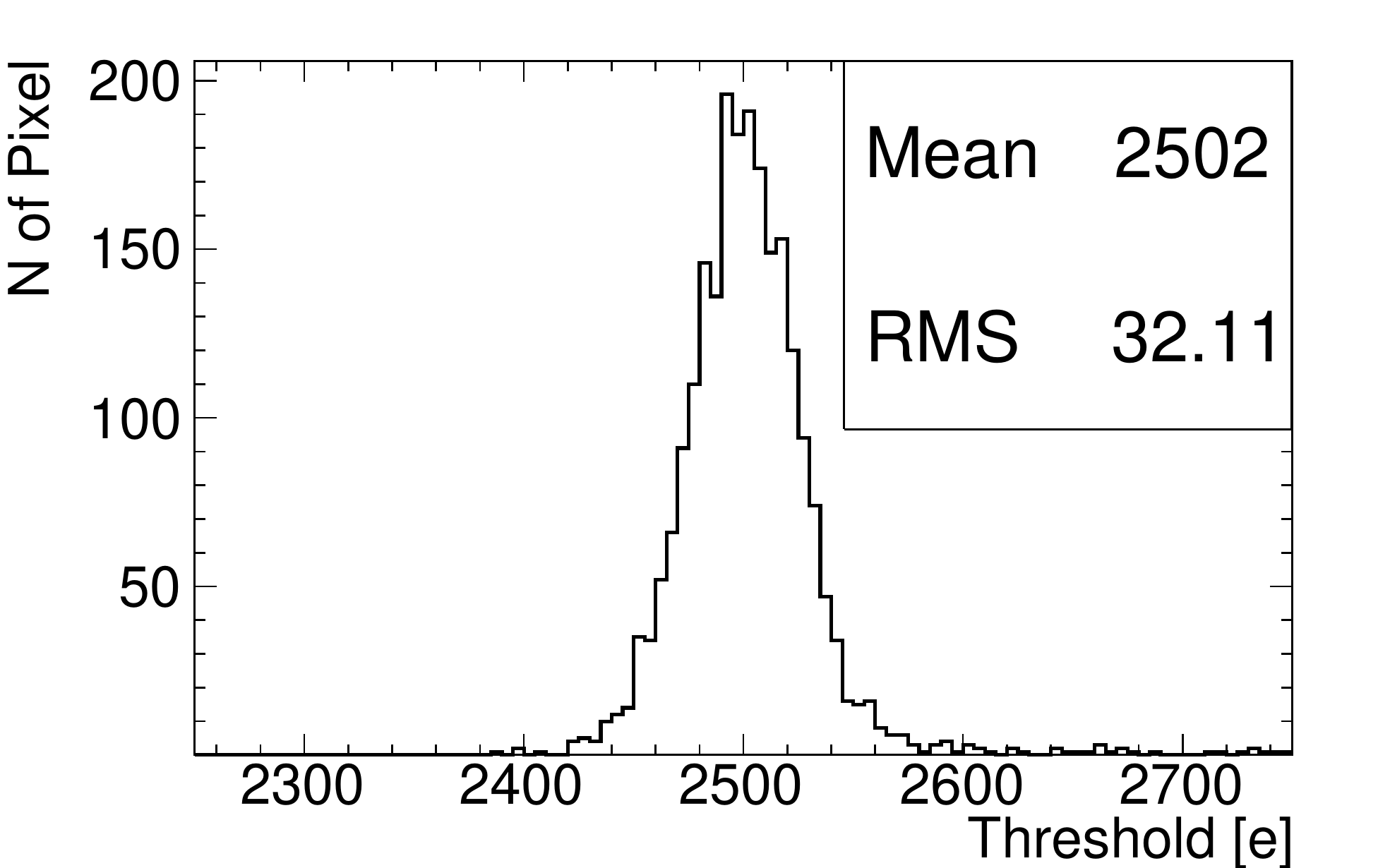}
\label{fig:SLID9_Thr2500_thr}
}
\subfigure[]{
\includegraphics[scale=0.35]{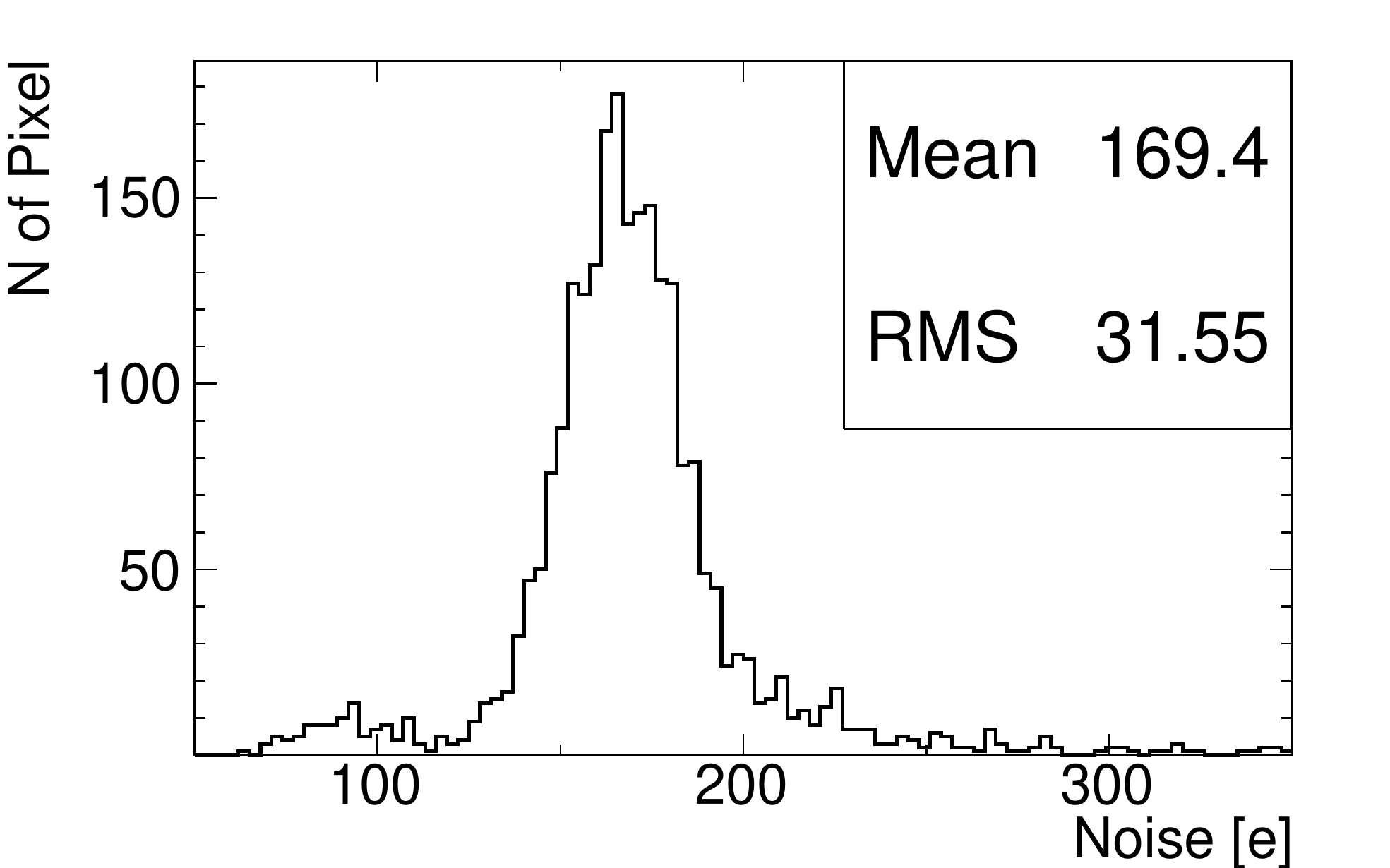}
\label{fig:SLID9_Thr2500_noise}
}
\caption{Tuned (a) threshold and (b) noise for the neutron irradiated sample.}
\label{fig:SLID9_Thr2500_thr_noise}
\end{figure}

One SCM was irradiated to a fluence of $2\cdot10^{15}$\,\neqcm{} using reactor neutrons at the Jo\v{z}ef-Stefan-Institut in Ljubljana \cite{Snoj2011136}. All measurements for the irradiated sample were conducted in a climate chamber at $-50$\,$^{\circ}$C. In Fig.\,\ref{fig:SLID9_Thr2500_thr_noise} the threshold tuning and corresponding noise for this SCM is shown. 
The mean threshold (cf. Fig.\,\ref{fig:SLID9_Thr2500_thr}) could be tuned as low as $2.5$\,ke with a dispersion of $32$\,e over the sensor. The corresponding noise is $169$\,e with a dispersion of $32$\,e over the chip.
                                                                                       
\begin{figure}[h!t]
\centering
\subfigure[]{
\includegraphics[scale=0.35]{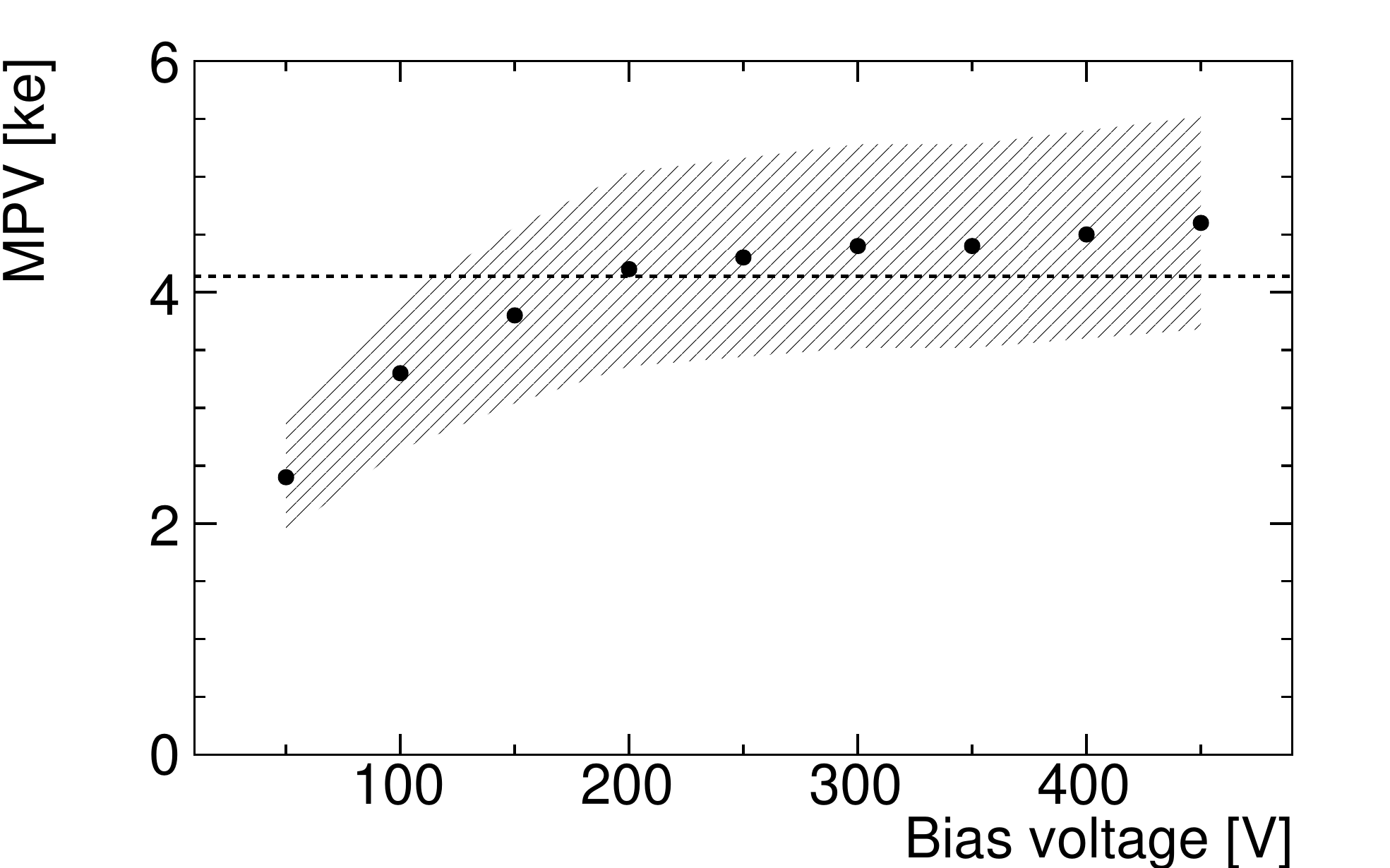}
\label{fig:SLID9irrad_Qvsbias}
}
\subfigure[]{
\includegraphics[scale=0.35]{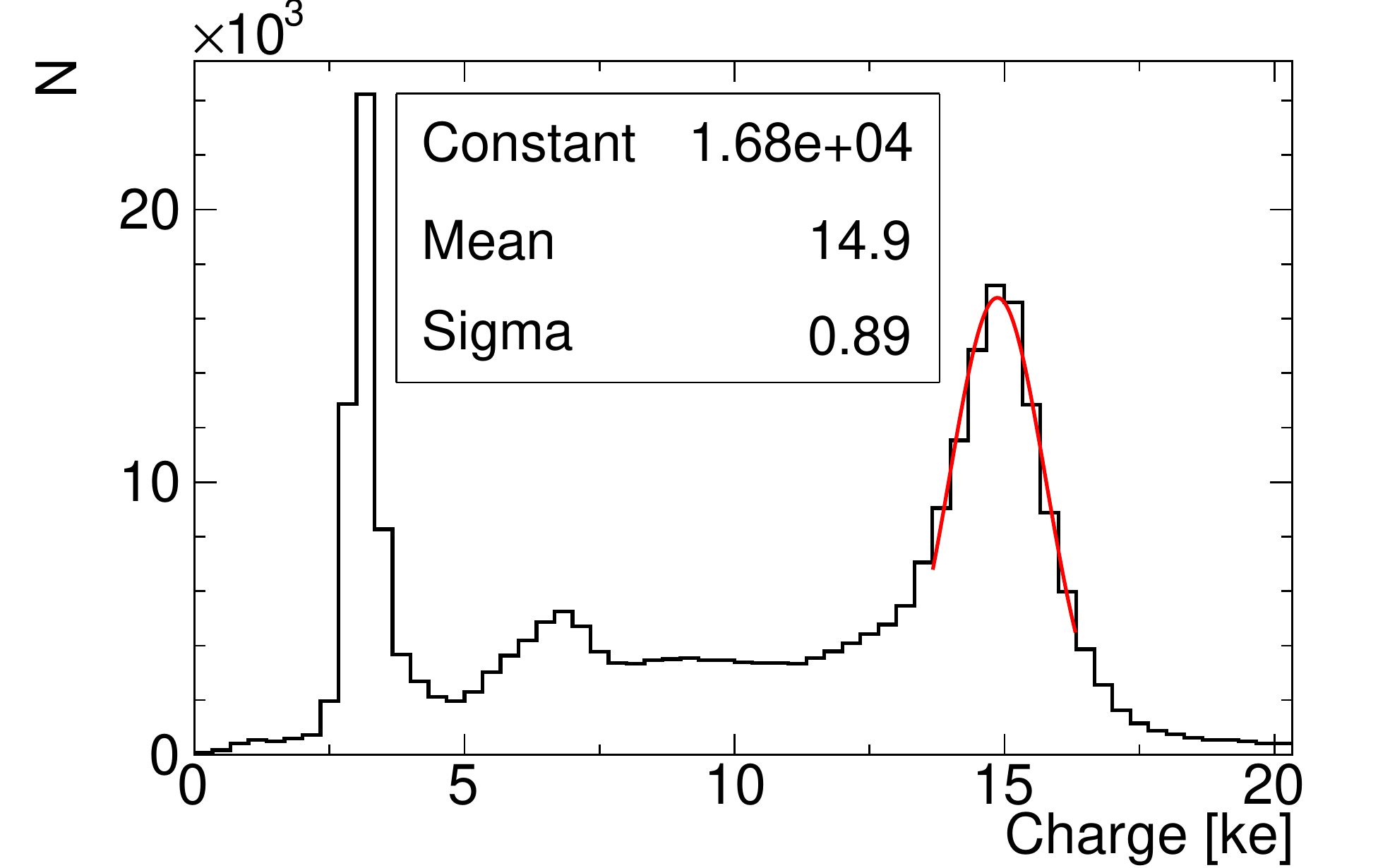}
\label{fig:SLID9irrad_Am241}
}
\caption{(a) MPV versus applied bias voltage for the neutron irradiated sample. The uncertainty band accounts for the overall time-over-threshold to charge calibration. The dotted line indicates the MPV before irradiation. (b) \Am{} spectrum at 300\,V.}
\label{fig:SLID9irrad}
\end{figure}

\Sr{} and an \Am{} source measurements are shown in Fig.\,\ref{fig:SLID9irrad}. In Fig.\,\ref{fig:SLID9irrad_Qvsbias} the dependence of the MPV, derived from \Sr{} measurements,  is shown versus the applied bias voltage. The uncertainty band is dominated by the calibration uncertainty, which is fully correlated over all measurements. Up to 200\,V a rise in the collected charge is observed which corresponds to an increase of the depleted volume. The bias voltage corresponding to the charge saturation value lies between the depletion voltages derived from charge collection efficiency measurements with strip devices irradiated to fluences of $10^{15}$\,\neqcm{} (150\,V) and $3\cdot10^{15}$\,\neqcm{} (250\,V). For comparison the MPV before irradiation at a bias voltage of 55\,V is indicated by a dotted line. The uncertainty band, which is almost fully correlated to the uncertainty after irradiation is not shown. In Fig.\,\ref{fig:SLID9irrad_Am241} an \Am{} measurement at a bias voltage of 300\,V is shown. This voltage corresponds to a comparable over-depletion as in the measurement before irradiation. The 59\,keV $\gamma$-line is fitted with a Gaussian and lies at $(14.9\pm3.0)$\,ke. The contributions of noise and of the $\gamma$-lines below 22\,keV are more pronounced than before irradiation for collected charges lower than 5\,ke given the reduced threshold and the slightly higher energy scale observed after irradiation when the sensor is over-depleted. Overall the performance before and after irradiation to a fluence of $2\cdot10^{15}$\,\neqcm{} using reactor neutrons is comparable in terms of charge collection.

Due to an imperfect opening of the Benzocyclobutene (BCB) passivation layer a fraction of channels is not connected with an increasing tendency towards the centre of the initial sensor wafer. The number of not connected channels did not change after irradiation and several thermal cylcles. This is an indication that SLID interconnections are radiation hard and withstand thermal cycles.

\subsection{Inter-Chip Vias}
Vertical signal transport to the backside of the read-out chip is permitted via ICVs. They are produced in a two-step process. First the via is etched, using the Bosch process \cite{boschpatent}, followed by the filling with tungsten to allow signal transport through the via. The ICV process of EMFT offers high aspect ratios up to 8:1, which are important to eventually transfer the process on a pixel-by-pixel level.
For this project FE-I3 chips are used, where several vias are etched into each wire-bond-pad. To insulate the pads from each other an additional trench is etched. Initial etching trials to depths between 50\,$\mu$m and 60\,$\mu$m were carried out to optimize the trench and ICV dimensions to achieve comparable etching depths, given their different geometries. The next step is to thin the chip wafer from the initial thickness of 650\,$\mu$m and form the metal contact pads on the back side. The R\&D will then continue with the SLID interconnection to thin sensors.

\section{Conclusions}
Measurements before and after irradiation of first demonstrator SCMs for a new module concept exploiting SLID interconnections were discussed. The modules show very good performance in terms of charge collection behaviour, both before and after irradiation. No changes in the SLID connection efficiency were observed after repeated thermal cycles as well as after high radiation levels. 

\acknowledgments
The authors thank I.~Mandic (Jo\v{z}ef-Stefan-Institut) for the sensor irradiation. The irradiation was supported by AIDA WP7. This work has been partially performed in the framework of the RD50 Collaboration.

\end{document}